%% file: main.tex
\documentclass[reprint,raggedbottom,aps,prb,amsfonts,amsmath,amssymb,floatfix,superscriptaddress]{revtex4-2}

\usepackage{graphicx,hyperref} 
\usepackage[all]{hypcap} 
\hypersetup{colorlinks=true,linkcolor=blue,citecolor=blue,urlcolor=blue}

\input{math_commands.tex}

\begin{document}

\title{Nonequilibrium free-energy calculation of solids using LAMMPS}
\author{Rodrigo Freitas}
\affiliation{Department of Materials Science and Engineering, University of California, Berkeley, CA 94720, USA} 
\affiliation{Instituto de Física ``Gleb Wataghin'', Universidade Estadual de Campinas, UNICAMP, Campinas, São Paulo 13083-859, Brazil} 
\email{rodrigof@berkeley.edu}
\author{Mark Asta}
\affiliation{Department of Materials Science and Engineering, University of California, Berkeley, CA 94720, USA} 
\author{Maurice de Koning}
\affiliation{Instituto de Física ``Gleb Wataghin'', Universidade Estadual de Campinas, UNICAMP, Campinas, São Paulo 13083-859, Brazil} 
\date{\today}

\begin{abstract}
  This article describes nonequilibrium techniques for the calculation of free energies of solids using molecular dynamics (MD) simulations. These methods provide an alternative to standard equilibrium thermodynamic integration methods and often present superior efficiency. Here we describe the implementation in the LAMMPS (Large-scale Atomic/Molecular Massively Parallel Simulator) code of two specific nonequilibrium processes that allow the calculation of the free-energy difference between two different system Hamiltonians as well as the free-energy temperature dependence of a given Hamiltonian, respectively. The theory behind the methods is summarized, and we describe (including fragments of LAMMPS scripts) how the process parameters should be selected to obtain the best-possible efficiency in the calculations of free energies using nonequilibrium MD simulations. As an example of the application of the methods we present results related to polymorphic transitions for a classical potential model of iron.
\end{abstract}

\maketitle
 
\section{Introduction}
 \label{sec:introduction}
 \input{sec_introduction}

\section{Nonequilibrium free-energy estimation}
  \label{sec:theory}
  \input{sec_theory}

\section{Calculation of bulk free energy}
  \label{sec:bulk}
  \input{sec_bulk}

\section{Iron polymorphism}
  \label{sec:polymorphism}
  \input{sec_polymorphism}

\section{Summary and discussion}
  \label{sec:summary}
  \input{sec_summary}

\section*{Acknowledgments}
  This work was supported by the FAPESP grant 2010/13902-4. MK acknowledges support from FAPESP and FAPESP/CEPID 2013/00293-7. The research of RF and MA at UC Berkeley were supported by the US National Science Foundation (Grant No. DMR-1105409).

\bibliography{bibliography}

\end{document}

%% file: math_commands.tex
\renewcommand{\v}[1]{\mathbf{#1}}        
\newcommand{\gv}[1]{\mbox{\boldmath$ #1 $}}

\renewcommand{\d}[0]{\ensuremath{\operatorname{d}\!}}
 
\newcommand{\td}[2]{\frac{\d #1}{\d #2}}        
        
\newcommand{\pd}[2]{\frac{\partial #1}{\partial #2}}

\let\f=\frac

\renewcommand{\l}[0]{\left}
\renewcommand{\r}[0]{\right}
\newcommand{\fp}[2]{\l( \f{#1}{#2} \r)}

\newcommand{\tf}[2]{\tfrac{#1}{#2}}

\newcommand{\avg}[1]{\l< #1 \r>}

\newcommand{\ie}[0]{\textit{i.e.}}
\newcommand{\eg}[0]{\textit{e.g.}}

\renewcommand{\t}[1]{\text{#1}}

\renewcommand{\and}[0]{\qquad \t{and} \qquad}

\newcommand{\Ang}[0]{\, \mathring{\mathrm{A}}}
\newcommand{\eV}[0]{\, \mathrm{eV}}

\newcommand{\kB}[0]{k_\t{B}}
\newcommand{\K}[0]{\, \t{K}}
\newcommand{\ns}[0]{\, \t{ns}}
\newcommand{\fs}[0]{\, \t{fs}}

\newcommand{\imp}[0]{\qquad \Rightarrow \qquad}


%% file: sec_introduction.tex
The calculation of free energies and derivative thermodynamic quantities for condensed-phase systems is a common and widespread application of atomistic simulation techniques. The continuous improvement of methods and the steady increase in computational power have enabled the efficient and reliable computation of free energies for complex systems of interest in materials science, including extended defects and interfaces. Free-energy calculation methods have benefited substantially with the introduction of nonequilibrium approaches such as the adiabatic switching~\cite{adiabatic_switching} (AS) method, which allow the calculations to be performed along explicitly time-dependent processes that can lead to significant efficiency gains compared to standard equilibrium methods. Moreover, with the derivation of Jarzynski's equality the connection between equilibrium free-energy differences and nonequilibrium processes has been placed on a firm theoretical basis \cite{jarzynski}.

In this paper we describe the implementation of state-of-the-art nonequilibrium techniques for calculating free energies of solids in the highly optimized LAMMPS~\cite{lammps} (Large-scale Atomic/Molecular Massively Parallel Simulator) molecular dynamics code. Specifically, we focus on two particular kinds of nonequilibrium routes which, respectively, allow the calculation of free-energy differences between systems described by different Hamiltonians as well as the temperature dependence of the free energy of a given system Hamiltonian from a single nonequilibrium simulation. We demonstrate their application to the calculation of free energies of different crystalline structures for a classical interatomic potential model of iron~\cite{eam_fe} and discuss extensions of the approaches to the calculations of interfacial free energies, which can be readily undertaken using the methods implemented in LAMMPS \cite{fix_fl}.

The paper has been organized as follows. In Sec.~\ref{sec:theory} we give a short and self-contained presentation of the theoretical framework underlying the nonequilibrium processes used to compute free-energy differences between equilibrium states. In Sec.~\ref{sec:bulk} we describe the implementation of these methods into LAMMPS, and discuss how to optimize the parameters for efficiency, using bcc iron as an example system. In Sec.~\ref{sec:polymorphism} we describe the application of the methods to the study of polymorphism in iron and we end with a summary and a discussion of the extension of the methodology to more complex crystalline systems as well as surfaces and interfaces in Sec.~\ref{sec:summary}.

%% file: sec_theory.tex
Standard equilibrium free-energy calculations are often based on thermodynamic integration (TI), \cite{kirkwood,ums} which is a general class of methods based on the construction of a sequence of equilibrium states on a path between two thermodynamic states of interest. The free-energy difference between two equilibrium states is then determined by computing ensemble averages of the relevant thermodynamic driving force for these states by means of a set of independent equilibrium simulations, followed by numerical integration. This approach embodies the thermodynamic equality between the free-energy difference between two equilibrium states and the reversible work $W_\t{rev}$ that is done along a quasistatic path that connects them. 

Nonequilibrium approaches, on the other hand, envision this path in terms of an explicitly time-dependent process. The rate at which this process is executed then determines the amount by which it deviates from a quasistatic path. Jarzynski's equality \cite{jarzynski} relates the work done along arbitrarily out-of-equilibrium processes, starting from an equilibrated initial state, to the free-energy difference $\Delta F$ between this state and the equilibrium state at the end of the process. In contrast to a quasistatic path, however, the work done in a nonequilibrium process is a stochastic variable that differs for each realization and simulations provide a means for sampling its distribution function. Jarzynski's equality then connects this distribution to the free-energy difference $\Delta F$ between the equilibrium states defined by the parameters at the two ends of the path through  
\begin{equation}
  \label{eq:jarz}
  \exp(-\beta \Delta F)=\overline{\exp(-\beta W_\t{irr})},
\end{equation} 
where $W_\t{irr}$ is the irreversible work done along such a nonequilibrium process and the overline denotes averaging over an ensemble of different realizations. While the equality is exact, its practical application in free-energy calculations is often limited due the exponential average in Eq.~\eqref{eq:jarz}, which may lead to very substantial statistical uncertainties in its evaluation \cite{Oberhofer2005}.

Another way to relate $\Delta F$ to the irreversible work distribution is by directly connecting its mean value to the true reversible work. One has
\begin{equation}
  \label{eq:dF2}
  \Delta F=W_\t{rev}=\overline{W_\t{irr}} - \overline{E_\t{diss}},
\end{equation}
where $\overline{E_\t{diss}}$ is the average dissipated heat generated for an ensemble of replicas of the nonequilibrium process. Because of the second law of thermodynamics we have $\overline{E_\t{diss}} \geq 0$, with the equality being valid only in the limit of an infinitely slow, quasistatic process. Instead of the exponential average in Eq.~\eqref{eq:jarz}, this relation involves a simple mean of $W_\t{irr}$ values. This significantly reduces the statistical uncertainties but the price to pay is the presence of the \textit{a priori} unknown systematic error in the form of the dissipated heat $\overline{E_\t{diss}}$. 

However, provided that the nonequilibrium process is sufficiently ``close'' to the ideally quasistatic process for linear response theory to be valid, it can be shown that this systematic error becomes the same for two processes that are carried out in opposite directions \cite{dekoning_2}. In other words, for a linear-response nonequilibrium process connecting states $1$ and $2$, we have $\overline{E^\t{diss}_{1 \to 2}}=\overline{E^\t{diss}_{2 \to 1}}$, from which it follows
\begin{eqnarray}
  \Delta F 
  & \equiv & F_2-F_1 \nonumber \\
  & \equiv & \tf{1}{2} \Big[ W^\t{rev}_{1\to2} - W^\t{rev}_{2\to1} \Big] \nonumber \\
  & = & \tf{1}{2} \bigg\{ \l[\overline{W^\t{irr}_{1\to2}} - \overline{E^\t{diss}_{1\to2}}\r] - \l[\overline{W^\t{irr}_{2\to1}} -\overline{E^\t{diss}_{1\to2}}\r] \bigg\} \nonumber \\
  & = & \tf{1}{2} \l[\overline{W^\t{irr}_{1\to2}} - \overline{W^\t{irr}_{2\to1}}\r], 
\label{eq:W12}
\end{eqnarray}
eliminating the systematic error by combining the results of the forward and backward processes. Similarly, the magnitude of the dissipation in this regime can be estimated by
\begin{equation}
  \label{eq:Ediss}
  \overline{E^\t{diss}_{1\to2}} =
  \overline{E^\t{diss}_{2\to1}} =
  \tfrac{1}{2} \l[\overline{W^\t{irr}_{1\to2}} + \overline{W^\t{irr}_{2\to1}}\r].
\end{equation} 

One of the challenges of standard equilibrium TI approaches is how to discretize the quasistatic path between the two states of interest. In addition to deciding on the number of states, it also requires choosing how to distribute them over the path. Furthermore, every state requires a separate simulation, each of which should allow sufficient equilibration as well as ensemble averaging time. 

In the nonequilibrium approach discussed above, on the other hand, the entire process is sampled during a single simulation and the closeness to equilibrium can be systematically assessed by monitoring the convergence of Eq.~\eqref{eq:W12} as a function of process rate. The characteristic that a desired free-energy value can be estimated from a few relatively short simulations and that its convergence can be systematically verified render it an attractive alternative to the standard equilibrium TI methodology and often gives substantially improved efficiency.

\subsection{Nonequilibrium free-energy differences for a parameter-dependent Hamiltonian}
  Consider a system of $N$ particles confined to a volume $V$ and in thermal equilibrium with a heat reservoir at temperature $T$. The Hamiltonian of the system is given by $H(\gv{\Gamma}, \lambda)$, where $\gv{\Gamma} = (\v{r}, \v{p})$ is a point in the phase space of the particles in the system (atoms in the present context) and $\lambda$ is a parameter, a specific example of which will be given below. The Helmholtz free energy of this system, for a particular value of $\lambda$, is $F(N, V, T; \lambda) = -\kB T \ln Z(N, V, T; \lambda)$ where $\kB$ is the Boltzmann constant and $Z$ is the system's canonical partition function given as an integral over the entire phase-space volume:
  \begin{equation}
    \label{eq:Z}
    Z(N, V, T; \lambda) = \int \f{\d{\gv{\Gamma}}}{h^{3N}} \exp \big[ -\beta H(\gv{\Gamma}, \lambda) \big],
  \end{equation}
  where $\beta = 1/\kB T$, and $h$ is Planck's constant. 
  
  Let us consider the problem of calculating the free-energy difference between two thermodynamic states characterized by different values of the parameter $\lambda$ (namely $\lambda_\t{i}$ and $\lambda_\t{f}$), \ie, our goal is to calculate $\Delta F(N, V, T) \equiv F(N, V, T; \lambda_\t{f}) - F(N, V, T; \lambda_i)$. From now on we will omit the dependence on $N$, $V$, and $T$ when their values are clear from the context.
  
  The desired free-energy difference can be computed by finding
  the derivative of the $F$ with respect to $\lambda$,
  \begin{equation}
    \label{eq:dFdL}
    \pd{F}{\lambda} = \f{1}{Z} \int \f{\d{\gv{\Gamma}}}{h^{3N}} \; \pd{H}{\lambda} \exp \big[ -\beta H(\gv{\Gamma}, \lambda) \big]
                  = \avg{\pd{H}{\lambda}}_\lambda,
  \end{equation}
  where $\avg{\ldots}_\lambda$ is the canonical ensemble average for a specific value of the parameter $\lambda$, and integrate it to obtain
  \begin{eqnarray}
    \Delta F = F(\lambda_\t{f}) - F(\lambda_\t{i}) & = & \int_{\lambda_\t{i}}^{\lambda_\t{f}} \d{\lambda} \avg{\pd{H}{\lambda}}_\lambda \label{eq:TI} \\
    & \equiv & W^\t{rev}_{\t{i}\to \t{f}} \nonumber .
  \end{eqnarray}
  The integral on the right-hand side can be interpreted as the reversible work along a quasistatic process between the equilibrium states with the two $\lambda$--values of interest. In the equilibrium TI approach, this integral is discretized on a grid of $\lambda$--values and for each value a separate equilibration simulation is executed.
   
  In the nonequilibrium approach the integral in Eq.~\eqref{eq:TI} is estimated in terms of the irreversible work done along a single simulation in which $\lambda = \lambda(t)$ is explicitly time dependent and varied from $\lambda_\t{i}$ to $\lambda_\t{f}$ in a switching time $t_\t{s}$,
  \begin{equation}
    \label{eq:w_irr}
    W_{\t{i} \to\t{f}}^{\t{irr}} = \int_0^{t_\t{s}} \d{t} \td{\lambda}{t} \l(\pd{H}{\lambda}\r)_{\gv{\Gamma}(t)} ,
  \end{equation} 
  where $\gv{\Gamma}(t)$ represents the phase-space trajectory of the system along the process. In practice, the integral in Eq.~\eqref{eq:w_irr} is evaluated in terms of the sum
  \begin{equation}
    W_{\t{i}\to\t{f}}^\t{irr}=\sum_{k=0}^{N-1} \Delta \lambda_k \l(\pd{H}{\lambda}\r)_{\gv{\Gamma}(k\Delta t)} ,
  \end{equation}
  where $\Delta t$ is the MD time step, $k$ is the time step number, $N$ is the total number of time steps in which $\lambda$ varies from $\lambda_\t{i}$ to $\lambda_\t{f}$ and $\Delta \lambda_k \equiv \lambda_{k+1} - \lambda_k$ is the discretization step of the switching parameter $\lambda$ at time step $k$.
  
  As mentioned previously, due to the nonequilibrium character of the process the average value of several realizations is subject to a systematic error due to the dissipated heat. But if the process is sufficiently slow for linear response theory to be accurate, it can be eliminated by combining the results of forward and backward switching processes as described by Eqs. (\ref{eq:W12}) and (\ref{eq:Ediss}). 
  
  Finally, it is important to emphasize that, before the nonequilibrium process is initiated, the system should be equilibrated at either $\lambda_\t{i}$ or $\lambda_\t{f}$, depending on the sense of the switching process~\cite{jarzynski}.
  
  Next we discuss the application of two specific thermodynamic paths that allow the calculation of the free-energy difference between two systems described by different Hamiltonians as well as the computation of the temperature variation of the free energy for a given system Hamiltonian.

\subsection{Free-energy difference between two systems: Frenkel--Ladd path}
  We choose the parametrical Hamiltonian $H(\lambda)$ to be of the particular form
  \begin{equation}
    \label{eq:interp} 
    H(\lambda)=\lambda H_\t{f} + (1-\lambda) H_\t{i},
  \end{equation}
  where $H_\t{i}$ and $H_\t{f}$ represent two different system Hamiltonians. Setting $\lambda_\t{i}=0$ and $\lambda_\t{f}=1$, respectively Eq.~\eqref{eq:TI} represents the free-energy difference between these systems, \ie,
  \begin{eqnarray}
    \nonumber
    \Delta F = F_\t{f} - F_\t{i} & = & \int_0^1 \d{\lambda} \avg{H_\t{f} -H_\t{i}}_\lambda \\
    \label{eq:TI2}
    & \equiv & W_{\t{i}\to\t{f}}^\t{rev},
  \end{eqnarray}
  and the corresponding forward irreversible work estimator is determined as
  \begin{equation}
    \label{eq:w_irr2}
    W_{\t{i}\to\t{f}}^\t{irr} = \int_0^{t_\t{s}} \d{t}\,\td{\lambda}{t} \Big[ H_\t{f}\big(\gv{\Gamma}(t)\big) - H_\t{i} \big(\gv{\Gamma}(t)\big) \Big].
  \end{equation}
  
  The Frenkel--Ladd (FL) path~\cite{frenkel_ladd} uses this concept for computing the absolute free energy of atomic solids. The initial Hamiltonian $H_\t{i}$ in Eq.~\eqref{eq:interp} is chosen to be that of the system of interest for which we wish to compute the free energy, usually of the form
  \begin{equation}
    \label{eq:H}
    H_\t{i} \equiv H_0 = \sum_{i=1}^N \f{\v{p}_i^2}{2m} + U(\v{r})
  \end{equation}
  where $U(\v{r})$ is some interaction potential and $m$ is the particle mass. We assume that at a certain temperature and volume we know that the stable phase of this system is a certain solid structure (\eg, face-centered cubic). As the second Hamiltonian in the path of Eq.~\eqref{eq:interp} we consider that of a system of noninteracting particles of mass $m$, each of which is attached to a lattice point by a 3-dimensional harmonic spring. The crystallographic lattice to which these particles are connected corresponds precisely to that of the equilibrium phase of interest of system $H_0$. The Hamiltonian of this harmonic reference system, known as an Einstein crystal, can be written as
  \begin{equation}
    \label{eq:einstein}
    H_\t{f} \equiv H_\t{E} = \sum_{i=1}^N \l[ \f{\v{p}_i^2}{2m} + \f{1}{2} m \omega^2 (\v{r}_i - \v{r}_i^0)^2 \r]
  \end{equation}
  where $\omega$ is the oscillator frequency and $\v{r}_i^0$ is the equilibrium position of particle $i$ in system $H_0$. Its Helmholtz free energy is known analytically, namely,
  \begin{equation}
    F_\t{E}(N, V, T) = 3 N \kB T \ln \fp{\hbar \omega}{\kB T}.
  \end{equation}
  
  By estimating the reversible work between these two states, combining the results of forward and backward switching processes (with $\lambda(t=0) = 0$ and $\lambda(t=t_\t{s}) = 1$ and the opposite for the forward and backward processes, respectively), as described previously, the free energy of interest can be estimated as 
  \begin{equation}
  \label{eq:fl}
    F_0(N, V, T) = F_\t{E}(N, V, T) + \tfrac{1}{2} \l(\overline{W_{\t{i}\to\t{f}}^\t{irr}}-\overline{W_{\t{f}\to\t{i}}^\t{irr}}\r).
  \end{equation}

\subsection{Temperature dependence of the free energy: the Reversible Scaling path}
  The Reversible Scaling (RS) path \cite{rs, rs_2} is a particular parametric form of the Hamiltonian $H(\lambda)$ for which each value of $\lambda$ corresponds to a particular temperature of a given system Hamiltonian $H_0$. In this way, applying the nonequilibrium free-energy approach allows the calculation of the temperature dependence of $F_0(N,V,T)$ from a single constant temperature simulation. In the remainder of this section we present the key equations of this method which was originally described in Ref.~\cite{rs}.

  Consider again the Hamiltonian of the system of interest $H_0$, given by Eq.~\eqref{eq:H}. Its free energy is given by
  \[
    F_0(T_0) = -\kB T \ln Q(T_0)  + 3N\kB T_0 \ln \Lambda (T_0)
  \]
  where $\Lambda(T_0) = (h^2 / 2\pi m\kB T_0)^{1/2}$ is the thermal de Broglie wavelength and
  \begin{equation}
    \label{eq:Q_1}
    Q(T_0) = \int \d^{3N}\v{r} \exp \Big[ - U(\v{r}) / \kB T_0 \Big] 
  \end{equation}
  is the configurational part of the partition function.

  We define the parametrical Hamiltonian $H(\lambda)$ by introducing a scaling factor $\lambda$ in the potential energy function of $H_0$ such that
  \begin{equation}
    \label{eq:H_rs}
    H(\lambda) = \sum_{i=1}^N \f{\v{p}_i^2}{2m} + \lambda U(\v{r}).
  \end{equation}
  The configurational part of the partition function of the system described by $H(\lambda)$ is
  \begin{equation}
    \label{eq:Q_2}
    \int \d^{3N}\v{r} \exp \Big[ - \lambda U(\v{r}) / \kB T_0 \Big] \equiv Q(T_0/\lambda).
  \end{equation}
  Because of the similarity between the configurational partition function of these two systems, Eqs.~\eqref{eq:Q_1} and \eqref{eq:Q_2}, one can show \cite{rs} that their free energies are related as
  \begin{equation}
    \label{eq:F0}
    F_0(T) = \f{1}{\lambda} F(T_0; \lambda) + \f{3}{2} N\kB T_0 \f{\ln \lambda}{\lambda}
  \end{equation}
  where $T \equiv T_0/\lambda$, and $F(T_0;\lambda)$ is the free energy of the system $H(\lambda)$ for a specific value of the parameter $\lambda$. Equation \eqref{eq:F0} shows that each value of $\lambda$ in the scaled Hamiltonian $H(\lambda)$ at temperature $T_0$ corresponds to the system described by $H_0$ at a temperature $T = T_0/\lambda$.

  With the scaled Hamiltonian of Eq.~\eqref{eq:H_rs} we can apply the nonequilibrium approach (Eq.~\eqref{eq:w_irr}) to estimate the irreversible work done when $\lambda(t)$ is varied from $\lambda(0) = 1$ to $\lambda(t_\t{s}) = \lambda_\t{f}$ during a single simulation performed at temperature $T_0$:
  \begin{equation}
    \label{eq:w_irr_rs}
    W_{1\to\lambda_\t{f}}^\t{irr} = \int_0^{t_\t{s}} \d{t}\,\td{\lambda}{t}  \, U\big(\gv{\Gamma}(t)\big).
  \end{equation}
  Then, temperature dependence of the free energy of the system described by $H_0$ is \cite{rs}
  \begin{equation}
    \label{eq:rs}
    F_0(T) = \f{F_0(T_0)}{\lambda} + \f{3}{2} N\kB T_0 \f{\ln \lambda}{\lambda} + \f{1}{2\lambda}\Big[ W^\t{irr}_{1\to\lambda} - W^\t{irr}_{\lambda\to1} \Big],
  \end{equation}
  from which it becomes clear that $F_0(T)$ can be calculated for all temperatures between $T_0$ and $T_0/\lambda_\t{f}$ using the irreversible work estimated along a single nonequilibrium simulation.
  
  Notice that in the application of Eq.~\eqref{eq:rs}, we start from knowledge of $F(T_0)$: the free energy at the reference temperature $T_0$ obtained previously using, for example, the Frenkel--Ladd path presented in the last section.

%% file: sec_bulk.tex
In this section we describe the implementation of the methods described above in the widely used Molecular Dynamics code LAMMPS \cite{lammps}. We demonstrate the application of the methods to the calculation of the temperature-dependent free energy of a body-centered-cubic solid described by an embedded-atom-method (EAM) \cite{eam} many-body interatomic potential model of iron developed by Meyer and Entel \cite{eam_fe}. Free energies are calculated at zero pressure for a range of temperatures between $100$ and $1600\K$; the Frenkel--Ladd path is used to compute the reference free energy at $T_0 = 100\K$, and RS to extend the calculations up to $T = 1600\K$. All simulations are performed using the nonequilibrium approach described in the previous section.

\subsection{Preparation for Frenkel--Ladd path\label{ssec:p_fl}}
  Before we run the simulations for the FL path some initial simulations must be performed. First, benchmark simulations were performed in a microcanonical (NVE) ensemble using different values for the timesteps ($\Delta t$) to check the energy conservation of the Velocity Verlet integrator for this EAM potential, leading to the choice of a value of $\Delta t = 1 \fs$ for the subsequent simulations. All the further simulations require a thermostat to keep the temperature constant. We have chosen the Langevin thermostat \cite{langevin} for this purpose.

  Although we will be computing the Gibbs free energy at $P = 0$, the FL simulations need to be run in an ensemble with fixed volume instead of fixed pressure. This is necessary because the Einstein crystal is a system of independent particles and does not allow the computation of pressure (via the virial pressure equation for example \cite{ums}). Consequently, the equilibrium volume (or lattice parameter) of the system must first be calculated when the system is in equilibrium at the desired temperature and pressure. To calculate the lattice parameter we ran a simulation at $T_0 = 100\K$ using a barostat \cite{parrinello_rahman} to keep the system at zero hydrostatic pressure. In LAMMPS this is done by using the fix commands:
  \begin{flushleft}
    \hspace{1cm} \texttt{fix 1 all nph aniso 0.0 0.0 1.0}\\
    \hspace{1cm} \texttt{fix 2 all langevin 100 100 0.1 999}
  \end{flushleft}

  Recalling that the FL path involves a reference system composed of harmonic oscillators, the value of the corresponding spring constant ($k$) in Eq.~\eqref{eq:einstein} must be specified. Even though, in principle, the FL path works for any value of $k$, in practice different values change its efficiency considerably. This happens because the energy dissipation during the switching process (Eq.~\eqref{eq:dF2}) is sensitive to how different the Hamiltonians $H_0$ and $H_\t{E}$ are during the switching. If we want to keep them as close as possible to each other when $H(\lambda)$ is changed according to Eq.~\eqref{eq:interp}, then it is desirable to choose the reference Einstein crystal to be as similar as possible to the system described by $H_0$. Thus, we choose a spring constant $k = m \omega^2$ which results in vibrational frequencies as close as possible to the characteristic vibrational spectrum of the solid of interest. One effective and popular strategy \cite{ums} is to measure the mean-squared displacement $\avg{(\Delta \v{r})^2}$ of atoms in the system (iron in this case), and use the equipartition theorem to obtain
  \begin{equation}
    \label{eq:msd}
    \f{1}{2} k \avg{\l(\Delta \v{r}\r)^2} = \f{3}{2} \kB T
    \imp
    k = \f{3 \kB T}{\avg{\l(\Delta \v{r}\r)^2}}.
  \end{equation}

  In LAMMPS the mean-squared displacement can be easily calculated using the compute \texttt{msd} command and saving the fourth value in the output array (also named \texttt{msd} here):
  \begin{flushleft}
    \hspace{2cm} \texttt{compute 1 all msd com yes}\\
    \hspace{2cm} \texttt{variable msd equal c\textunderscore 1[4]}
  \end{flushleft}
  Notice that it is important to compute the spring constant using a simulation with exactly the same size and temperature as those intended to be used for the Frenkel--Ladd simulation. The reason is that the chosen spring constant is strongly related to the phonon spectrum of the system, which is affected both by temperature and system size.

  For the results presented here we have chosen a cubic simulation box with $18\times 18 \times 18$ bcc unit cells, or $11,664$ atoms. The equilibrium lattice parameter obtained at $T_0 = 100\K$ was $a(T_0) = 2.8841(1) \Ang$ and the spring constant was $k(T_0) = 5.787(1) \eV/\Ang^2$. 

  A cautionary remark about the simulations at constant temperature is to avoid the so-called ``flying ice cube'' problem \cite{ice_cube}. This problem arises when the thermostat used in the Molecular Dynamics simulation is allowed to act on the center-of-mass degrees of freedom and this bias is not removed when measuring the system temperature. If the thermostat is allowed to act on the center-of-mass degrees of freedom the entire system can eventually obtain a total drift velocity which contributes to the kinetic energy -- and therefore to the measured temperature -- but does not effectively raise the real temperature of the system, which should be measured considering only the internal degrees of freedom. Therefore, for all simulations we keep the center of mass fixed and do not allow the thermostat to add energy to it. Although there is more than one way to achieve this in LAMMPS, an effective and precise manner for doing this for the simulations we consider is to: (1) compute the temperature using only internal degrees of freedom, (2) make sure no fix changes the center-of-mass position, (3) start with zero total drift velocity. For example, we can enforce the fix \texttt{nph} to expand and contract the simulation box around the initial center-of-mass coordinates so that it does not change it:
  \begin{flushleft}
    \texttt{variable xcm equal xcm(all,x)} \\
    \texttt{variable ycm equal xcm(all,y)} \\
    \texttt{variable zcm equal xcm(all,z)} \\
    \texttt{fix f1 all nph aniso 0.0 0.0 1.0 \&} \\
    \hspace{2.65cm} \texttt{fixedpoint \$\string{xcm\string} \$\string{ycm\string} \$\string{zcm\string} } \\
  \end{flushleft}
  We also force the barostat and thermostat to compute the temperature after excluding the center-of-mass contribution with:
  \begin{flushleft}
    \texttt{fix f2 all langevin 100.0 100.0 0.1 999 zero yes} \\
    \texttt{compute c1 all temp/com} \\
    \texttt{fix\textunderscore modify f1 temp c1} \\
    \texttt{fix\textunderscore modify f2 temp c1} 
  \end{flushleft}
  Notice how we have explicitly used the \texttt{zero yes} option for the Langevin thermostat so that it does not add a drift when thermostatting the system.

\subsection{Frenkel--Ladd path in LAMMPS\label{ssec:fl_lammps}}
  Now that we have the equilibrium lattice constant and the optimal spring constant at $T_0 = 100\K$ we can use the Frenkel--Ladd path to compute the free energy. This is a constant temperature and volume simulation which uses the fix \texttt{ti/spring} command that we have implemented recently in LAMMPS. This fix works by performing the time-dependent switching between the Hamiltonian of the EAM iron potential and the Einstein crystal, as given in Eq.~\eqref{eq:interp}. The fix syntax is
  \begin{center}
    \texttt{fix f\textunderscore ID g\textunderscore ID ti/spring k t\textunderscore s t\textunderscore eq [function n]}
  \end{center}
  where \texttt{f\textunderscore ID} is the fix ID name (used by other commands that refer to this fix or the quantities it computes), \texttt{g\textunderscore ID} is the group ID of the group of atoms this fix acts on, \texttt{ti/spring} is the fix name, \texttt{k} is the spring constant of the Einstein crystal used (in the units defined in the simulation script), \texttt{t\textunderscore s} is the switching time for the full switch between $H_0$ and $H_\t{E}$ (in number of timesteps), \texttt{t\textunderscore eq} is the number of timesteps the system is allowed to equilibrate before the switching procedure begins, and \texttt{[function n]} is an optional keyword used to select the time-dependence functional form of the coupling parameter $\lambda(\tau)$, where $\tau = \Delta t/t_\t{s}$ is the fraction of the total switching time elapsed.

  The \texttt{ti/spring} fix changes the system Hamiltonian and the $\lambda$ parameter in Eq.~\eqref{eq:interp} as follows:
  \begin{enumerate}
    \item For the first \texttt{t\textunderscore eq} timesteps after the fix command was declared the $\lambda$ value will be zero to allow the system to equilibrate using the $H_0$ Hamiltonian.
    \item After that, during the next \texttt{t\textunderscore s} timesteps the $\lambda$ value will change gradually from $\lambda = 0$ to $\lambda = 1$ so that at the end of \texttt{t\textunderscore s} steps the system effective Hamiltonian is $H_\t{E}$. $\lambda(\tau)$ varies with $\tau$ according to the chosen function. This is the ``forward'' part of the simulation.
    \item The value of $\lambda$ is then fixed at $\lambda = 1$ for \texttt{t\textunderscore eq} steps. This is to allow the system to reach equilibrium in the $H_\t{E}$ Hamiltonian.
    \item Now for \texttt{t\textunderscore s} steps $\lambda$ will change back from $\lambda = 1$ to $\lambda = 0$ according to the $\lambda(\tau)$ function chosen. This is the ``backward'' part of the simulation.
    \item From now on the system Hamiltonian is $H_0$ and the fix has no effect.
  \end{enumerate}

  Two functional forms for $\lambda(\tau)$ are possible. The first one is a simple linear form $\lambda(\tau) = \tau$ and is specified by the keyword \texttt{function 1}. The second option \cite{einstein_maurice} is
  \[
    \lambda(\tau) = \tau^5 \l( 70\tau^4 - 315\tau^3 + 540\tau^2 - 420\tau + 126 \r)
  \]
  and is chosen by the keyword \texttt{function 2}. This function was implemented because it has a vanishing slope at the end of the switching process, \ie, $\d{\lambda}/\d{\tau} \rightarrow 0$ as $\tau \rightarrow 0$ or $\tau \rightarrow 1$. Is has been shown \cite{einstein_maurice} that functional forms of this type result in a less dissipative switching process.

  Because the MD simulations are performed with the constraint of fixed center of mass, we need to modify Eq.~\eqref{eq:fl} to account for this constraint. The first order approximation \cite{cm} to the contribution due to the fixed center of mass is that we should add to the right-hand side of Eq.~\eqref{eq:fl} the term
  \begin{equation}
    \label{eq:cm}
    \delta F_\t{CM} = \kB T \ln \l[ \f{N}{V} \fp{2\pi \kB T}{N m \omega^2}^{3/2} \r]
  \end{equation}
  where $N$ is the number of atoms and $V$ is the system total volume. Notice that this contribution vanishes in the thermodynamic limit as $\ln N / N$ per particle.

\begin{figure}[!ht]
    \centering
    \includegraphics[width=0.48\textwidth]{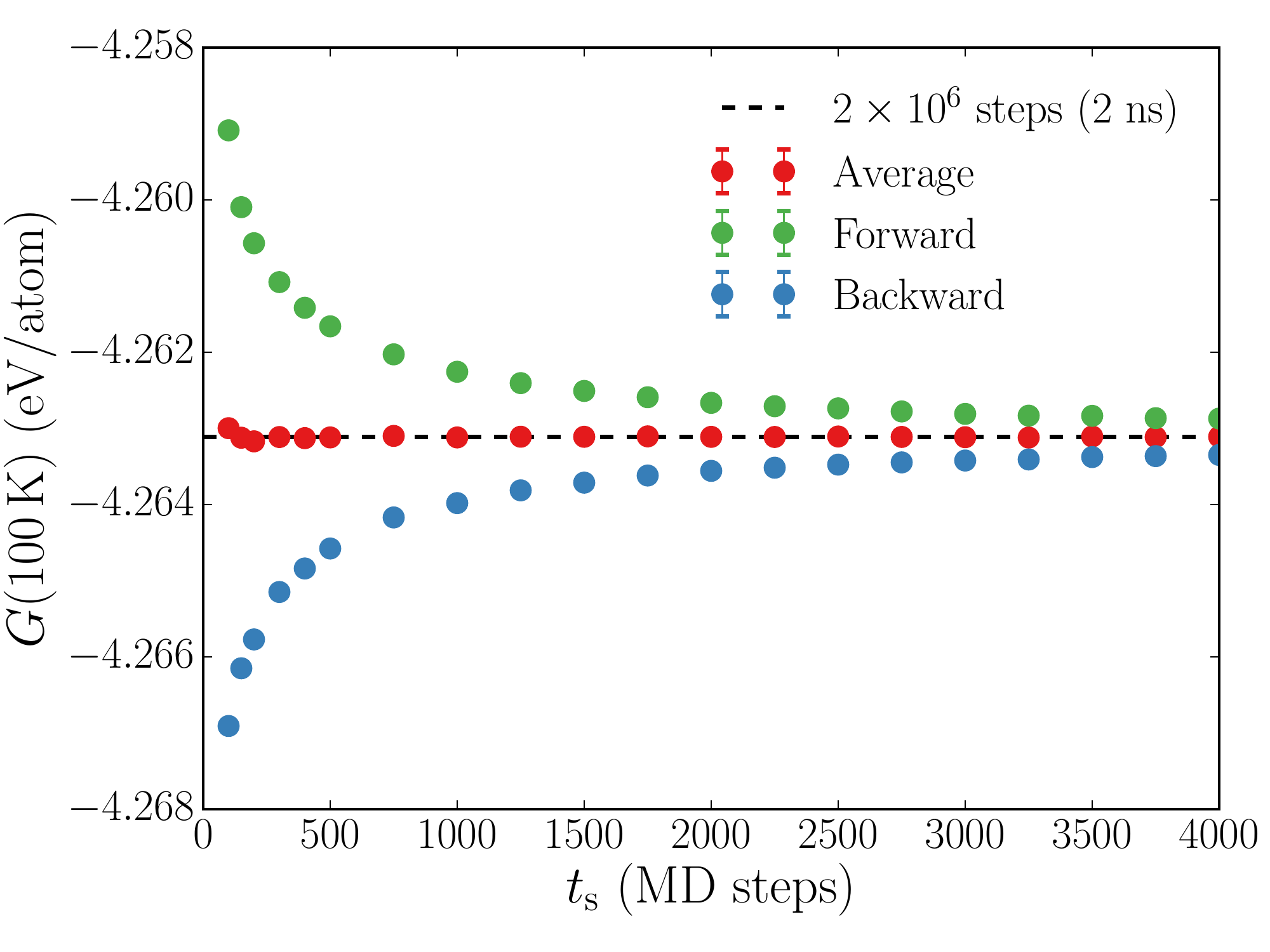}
    \caption{\label{fig:fl_convergence}Convergence of the nonequilibrium approach using a Frenkel--Ladd path to compute the free energy per atom of the bcc phase of pure iron. The systematic error introduced by the dissipation inherent of the nonequilibrium approach is easily eliminated by averaging the results of the forward and backward switching directions.}
  \end{figure}

  We have performed the forward and backward Frenkel--Ladd switching procedures in a cubic $18 \times 18 \times 18$ simulation cell ($11,664$ atoms) at $T_0 = 100\K$ using a spring constant of $k = 5.787 \eV/\Ang^2$. We have chosen an equilibration time of $t_\t{eq} = 0.1 \ns$ before starting the nonequilibrium switching along the FL path. Ten independent switching realizations were performed so that we could obtain an estimate for the statistical error. Equation \eqref{eq:fl} was used to compute the free energy and the correction for the fixed center of mass (Eq.~\eqref{eq:cm}) was included. Figure \ref{fig:fl_convergence} shows how the computed free energy converges with increasing the switching time $t_\t{s}$. In particular we see that the combination of the forward and backward paths (Eq.~\eqref{eq:fl}) is extremely efficient in eliminating the systematic error of the nonequilibrium approach. With a switching time as short as $300$ MD steps the unbiased estimate given by Eq.~\eqref{eq:fl} differs from a switching using $2\times10^6$ MD steps by about $10^{-2}\, \t{meV/atom}$. Using $t_\t{s} = 2\ns$  we arrived at a free energy of $G(T_0=100\K) = -4.2631147(1) \eV / \t{atom}$. Also, the statistical fluctuations associated with the stochastic nature of the irreversible work estimators are so small that the corresponding error bars are smaller than the symbols used in the plot.

\subsection{Reversible Scaling in LAMMPS}
  We now use the value of $G(T_0)$ at zero pressure obtained with the Frenkel--Ladd path as the reference point for the RS integration (Eq.~\eqref{eq:rs}). This simulation is run with a thermostat to keep the temperature constant at $T_0$ and a barostat to maintain zero hydrostatic pressure. Once again we choose the Langevin thermostat \cite{langevin} and the fix \texttt{nph} as barostat, taking the precautions discussed in Sec.~\ref{ssec:p_fl} to enforce a fixed center of mass. During the switching the system Hamiltonian has a scaled potential energy $\lambda U(\v{r})$ and $\lambda$ varies from $\lambda_\t{i} = 1$ to $\lambda_\t{f}$ during the simulation. According to Eq.~\eqref{eq:rs} this gives us the Gibbs free-energy temperature dependence from $T_0$ to $T_0/\lambda_\t{f}$ at $P = 0$. 

  The scaling of the interatomic potential was obtained in LAMMPS using the fix \texttt{adapt}. This command allow us to easily change simulation parameters over time, in our case the parameter will be $\lambda$. To introduce $\lambda$ as a variable that multiplies the potential energy function we use:
  \begin{flushleft} 
    \hspace{0.1cm} \texttt{fix f3 all adapt 1 pair eam scale * * v\textunderscore lambda}.
  \end{flushleft} 
  The syntax of this command is as follows: \texttt{f3} is the fix ID, \texttt{all} is the group ID of the atoms which the fix acts on, \texttt{adapt} is the name of the fix, \texttt{1} is to indicate the fix will be called every timestep, \texttt{pair eam scale} is the option to define that the parameter which will be changing over time will multiply (scale) the interatomic potential of type EAM, \texttt{* *} indicates that the interaction between all atoms types pairs will be affected, and finally \texttt{v\textunderscore lambda} identifies that the time-dependent variable which scales the interatomic potential is \texttt{lambda}.

  At this point it is worth commenting on why the fix \texttt{adapt} is used instead of developing a simpler fix \texttt{ti/rs} with a syntax similar to the one presented in Sec.~\ref{ssec:fl_lammps} for \texttt{ti/spring}. During the RS path we keep the pressure constant with the Parrinello-Rahman \cite{parrinello_rahman,martyna} barostat given by fix \texttt{nph}. This barostat effectively changes the equations of motion to be integrated \cite{shinoda} and a new integrator algorithm \cite{tuckerman_npt} is applied when this fix is invoked. If we naively scaled the energy and forces by $\lambda$ during a determined step of this integrator the effect of the barostat degrees of freedom on the particles would be scaled as well. Therefore, the correct strategy to perform the scaling inside the code is to change the interatomic potential directly, which can only be accomplished by the fix \texttt{adapt}. Note that the Frenkel--Ladd path implementation (\texttt{ti/spring}) does not suffer this same shortfall since the thermostat (fix \texttt{langevin}) does not require a different integrator. Thus, declaring the fix \texttt{ti/spring} before the fix \texttt{langevin} is sufficient to ensure that the code performs the interpolation of the many-body interatomic potential (Eq.~\eqref{eq:interp}) before the system is thermostatted.

  The scaling parameter $\lambda(\tau)$ was varied during the simulation according to the function
  \[
    \lambda(\tau) = \f{1}{1+\tau\l(\lambda_\t{f}^{-1}-1\r)}.
  \]
  This specific function was used because it results in a constant rate of change of $T$ with $\tau$. While it is possible that other functional forms for $\lambda(t)$ lead to smaller dissipation, in practice this particularly simple form has been found to give quickly converging results for RS calculations and we use it for all such calculations reported in this paper. The functional form given above for the parameter $\lambda(\tau)$ is obtained in LAMMPS by defining a variable:
  \begin{flushleft} 
    \hspace{0.1cm} \texttt{variable lambda equal \&}\\
    \hspace{2.6cm} \texttt{ 1/(1+elapsed/\$\string{t\textunderscore s\string}*(1/\$\string{lf\string}-1))}
  \end{flushleft} 
  where \texttt{t\textunderscore s} is the switching time $t_\t{s}$ and \texttt{lf} is $\lambda_\t{f}$.

  One may ask, how to choose the reference temperature $T_0$ for the RS path? In principle the results obtained are independent of the value of $T_0$ chosen. However, low temperatures usually result in better precision and rapid convergence of the Frenkel--Ladd path. Of course, if the purpose is to compare the results of simulations to experimental measurements then the classical simulation results should be considered only for temperatures above the Debye temperature.

\begin{figure}[!ht]
    \centering
    \includegraphics[width=0.48\textwidth]{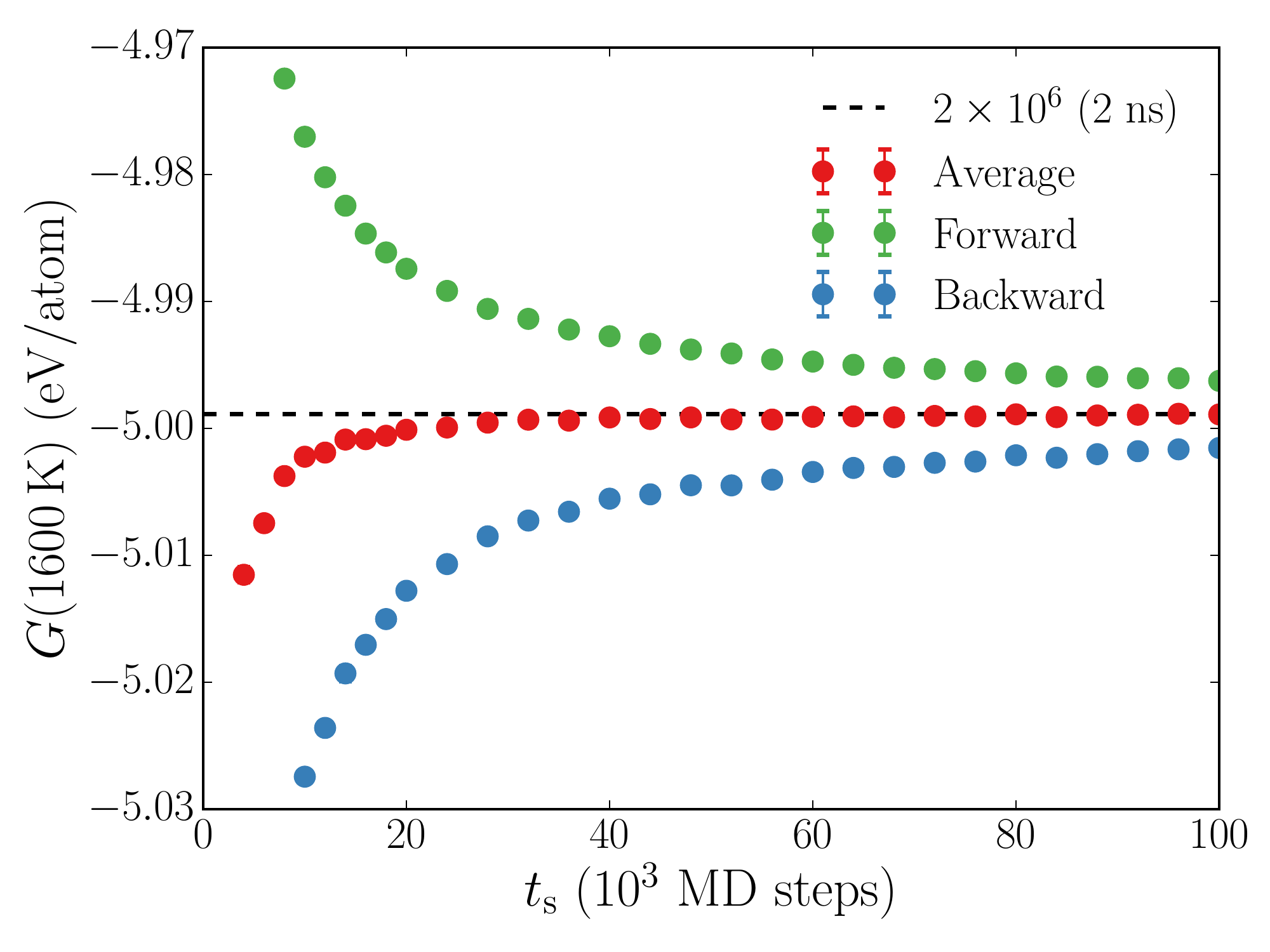}
    \caption{\label{fig:rs_convergence}Convergence of the nonequilibrium approach using a Reversible Scaling path to compute the free energy per atom of the bcc phase of pure iron. The reference point for the RS path was at $T_0 = 100 \K$ and $P = 0$, we show the free energy at the end of the RS path (at $1600\K$ and $P = 0$) where the effects of the energy dissipation are more relevant. The systematic error introduced by the dissipation inherent of the nonequilibrium approach is easily eliminated by averaging the results of the forward and backward switching directions.}
  \end{figure}

  The RS path chosen was such that $\lambda_\t{i} = 1$ and $\lambda_\t{f} = 0.0625$. According to Eq.~\eqref{eq:rs} this gives us the free-energy temperature dependence from $T_0 = 100\K$ to $T_0/\lambda_\t{f} = 1600\K$ at $P = 0$. Once again we equilibrate the system for $t_\t{eq} = 0.1\ns$ before starting the switching and performed ten independent forward and backward simulations to estimate the statistical error and the dissipation (according to Eq.~\eqref{eq:Ediss}). In Fig.~\ref{fig:rs_convergence} we show the convergence of the free energy at $T_0/\lambda_\t{f} = 1600\K$, at the end of the RS path, where the cumulative dissipation effects are the largest. With a nonequilibrium switching as short as $4\times10^4$ MD steps we are able to reproduce the free energy curve from $100\K$ to $1600\K$ with a precision of $1\,\t{meV/atom}$.

\begin{figure}[!ht]
    \centering
    \includegraphics[width=0.48\textwidth]{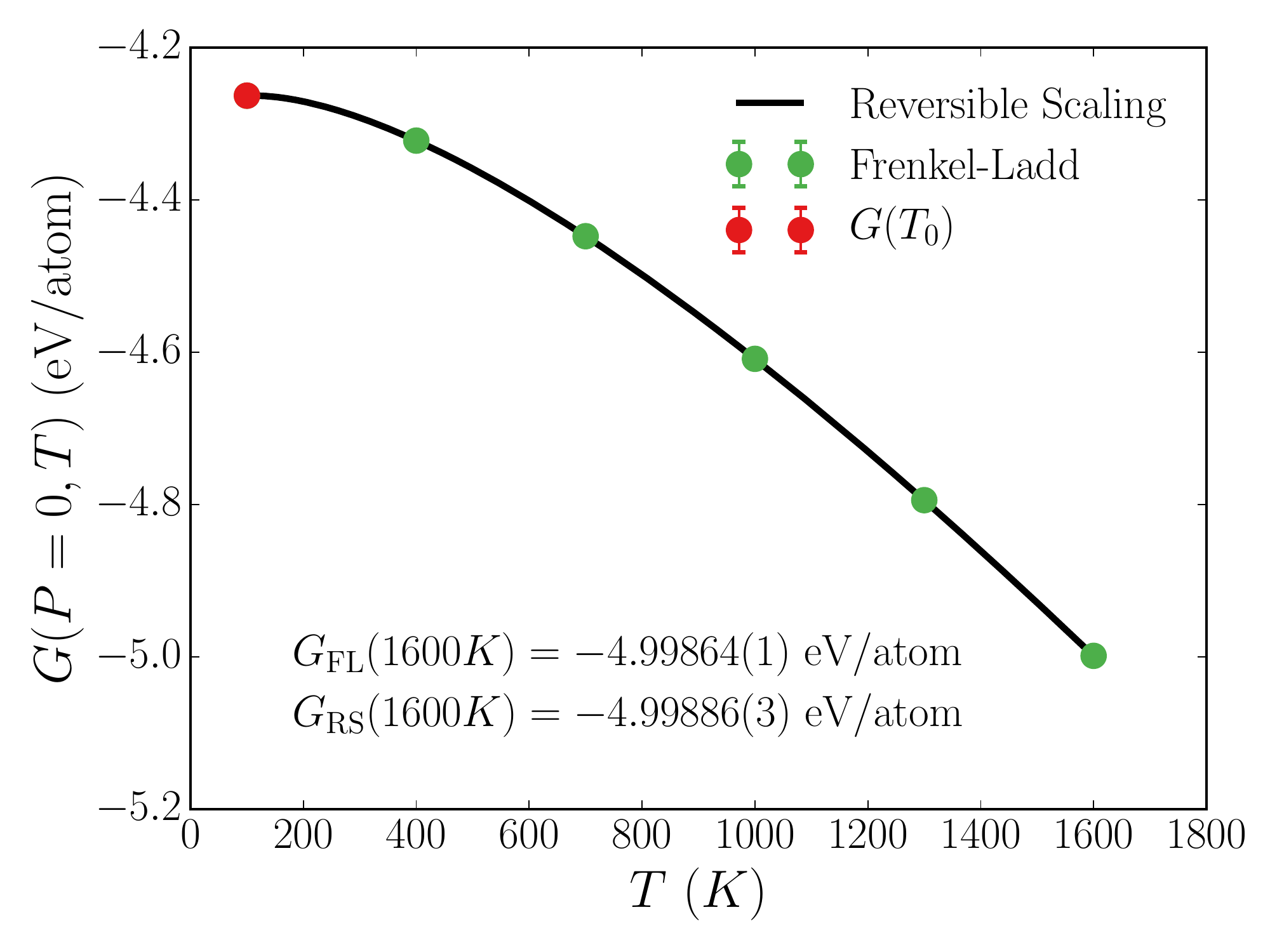}
    \caption{\label{fig:bcc} Free energy per atom of the bcc phase of pure iron at zero pressure. The reference point for the RS path was at $T_0 = 100 \K$ and $P = 0$. The FL calculations at other temperatures were used to verify the agreement against the RS result at high temperatures.}
  \end{figure}

  In Fig.~\ref{fig:bcc} we reproduce the temperature dependence curve for a switching time of $t_\t{s} = 2.0\ns$. We have also run additional Frenkel--Ladd simulations at $T = 400,\, 700,\, 1000,\, 1300,\, \t{and}\, 1600\K$ to check the agreement between both paths, which is shown to be excellent.

  We have chosen to compute the free-energy curve over a wide range of temperatures for the purpose of illustrating how the path works. This is not necessary and the RS path can be used more efficiently by traversing only a narrow temperature range near the region of interest, reducing dissipation and increasing the precision of the final result \cite{wei_cai}.

%% file: sec_polymorphism.tex
The formalism described in this article provides an efficient framework for computing the free energies of crystalline solids described by classical interatomic potential models. Such calculations are useful in many contexts in computational materials science. For example, a central issue in the modeling of crystalline solids is polymorphism, \ie, the thermodynamic stability of crystalline phases with different crystal structures as a function of temperature. Many materials used in engineering applications display polymorphism, such as iron which is bcc at low and high temperature and fcc at intermediate temperatures, or zirconia which transforms from a low-symmetry monoclinic structure at low $T$ to a cubic fluorite structure at high $T$. In many applications of atomistic simulations to the study of such materials, it is important that the interatomic potential model underlying the studies reproduce the stability of the desired phase or phases over a given temperature range. In other words, it is useful to characterize the relative stability of competing phases as a function of temperature as predicted by a given interatomic potential. We demonstrate the application of the nonequilibrium formalism to such calculations in this section, considering again the EAM potential for Fe developed in Ref.~\cite{eam_fe}. We note that this potential was not fit to reproduce experimentally-observed solid-state phase transitions in iron and indeed it is expected to be necessary to include explicitly magnetic degrees of freedom in the potential for this purpose \cite{meam_lee,abop_muller}. Thus, it is not expected for the potential to reproduce experimentally measured phase diagrams, and we use it here only to highlight the application of the nonequilibrium approach formalism to the study of polymorphism for a given interatomic potential.

\subsection{Free-energy size dependence}
  In this example we will consider the free energies of bcc, fcc and hcp crystal structures for the EAM potential of Fe given in Ref.~\cite{eam_fe}. Before presenting the free-energy results, we discuss system-size effects. All methods presented in this paper are designed to compute the vibrational free energy. Therefore, in addition to the center of mass term (Eq.~\eqref{eq:cm}), the size effects are due to the cutoff in the phonon spectrum introduced by the finite lattice size.

\begin{figure}[!ht]
    \centering
    \begin{flushleft}
      \large{(a)}
    \end{flushleft}
    \vspace{-0.4cm}
    \includegraphics[width=0.48\textwidth]{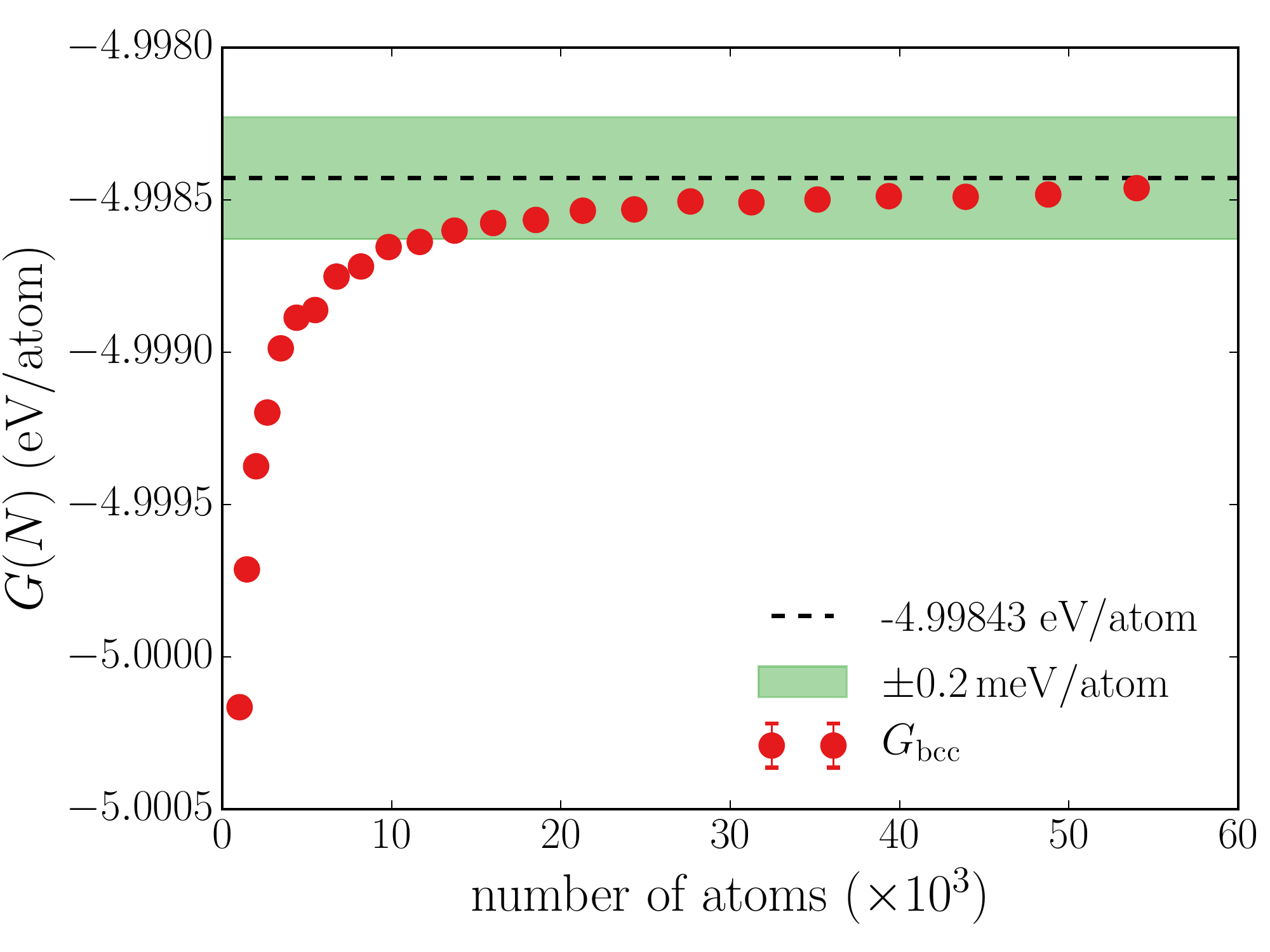}
    \vspace{-0.8cm}
    \begin{flushleft}
      \large{(b)}
    \end{flushleft}
    \vspace{-0.4cm}
    \includegraphics[width=0.48\textwidth]{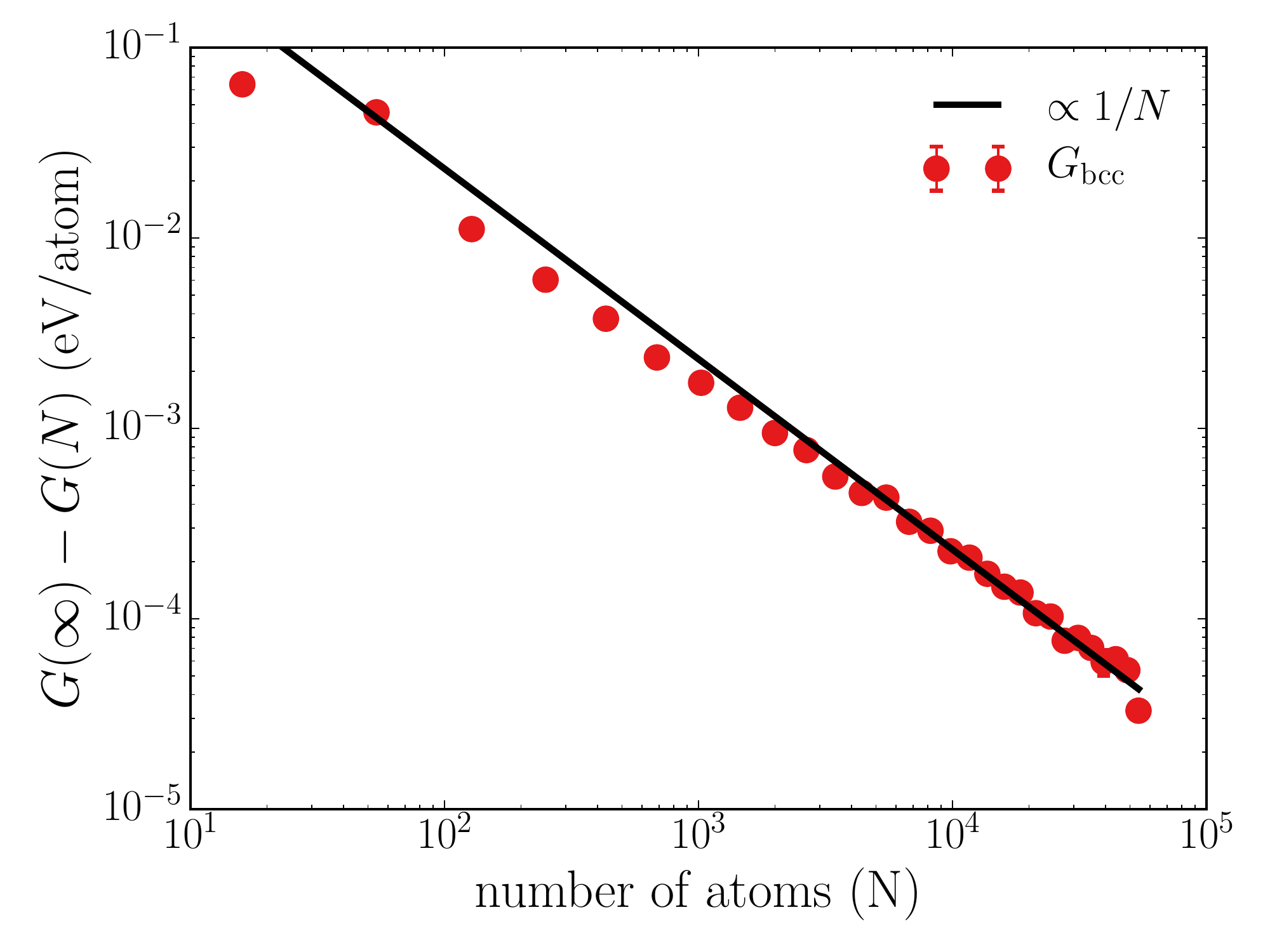}
    \caption{\label{fig:bcc_size_convergence}(a) Free-energy convergence with system size for the bcc structure at $1600\K$ and zero pressure. The chosen accuracy was within $\pm0.2\,\t{meV}/\t{atom}$ of the free energy in the thermodynamic limit. (b) We obtained an estimate for the free energy per atom in the thermodynamic limit, $G(\infty)$, by making an asymptotic analysis which has shown that the free energy, $G(N)$, converges with leading term $1/N$, in agreement with analytical calculations \cite{cm}.}
  \end{figure}

  Following the procedure described in Sec.~\ref{ssec:fl_lammps} we computed the free energy of the bcc phase at $1600\K$ for cubic simulation boxes with $M\times M\times M$ unit cells where $M = 1, 2, \ldots, 30$. The largest system considered had $N = 54,000$ atoms. The analysis of finite size effects in crystalline solids \cite{cm} has shown that, after considering the correction due to the fixed center of mass (Eq.~\eqref{eq:cm}), the free energy per atom converges with leading term $1/N$. The result for the convergence of the free energy of the bcc structure with the number of atoms is shown in Fig.~\ref{fig:bcc_size_convergence}(a), where the dashed line indicates an estimate of the thermodynamic limit, obtained from an asymptotic analysis of the free energy as a function of $1/N$ (Fig~\ref{fig:bcc_size_convergence}(b)). Based on similar simulations for the fcc and hcp structures we determined that, to converge the solid free energies to an accuracy of $0.2\,\t{meV/atom}$, the minimum system size was $18\times18\times18$ ($11,664$ atoms) for the bcc lattice, $14\times14\times14$ ($10,976$ atoms) for the fcc lattice, and $19\times11\times12$ ($10,032$ atoms) for the hcp lattice. Notice that because of the lack of cubic symmetry for the hcp structure the RS simulations have to be run with a barostat that controls the stress along each direction independently. This can be easily achieved in LAMMPS by selecting the \texttt{aniso} option in the fix \texttt{npt} command:
  \begin{flushleft}
    \hspace{1cm} \texttt{fix f1 all nph aniso 0.0 0.0 1.0}
  \end{flushleft}

\subsection{Results and discussion}
  The method for computing the free-energy curves of each phase is exactly the same as the method used in Sec.~\ref{sec:bulk} to compute the free energy of the bcc phase, shown in Fig.~\ref{fig:bcc}.   

\begin{figure}[!ht]
    \centering
    \includegraphics[width=0.48\textwidth]{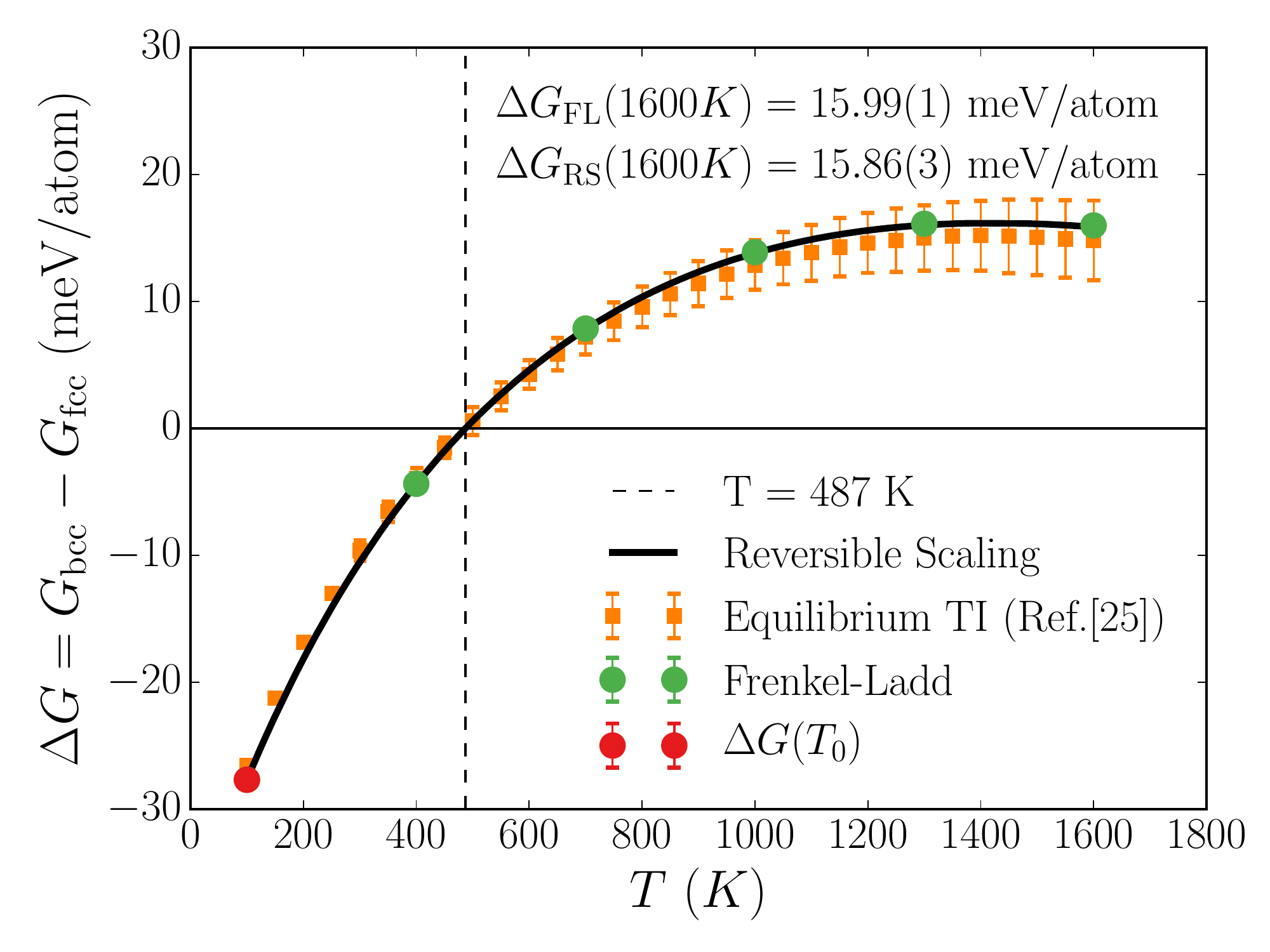}
    \caption{\label{fig:bcc_fcc}Free-energy difference between the bcc and fcc phases of iron. The bcc phase is stable below $487\,\t{K}$ while fcc is stable above it. The nonequilibrium approach discussed in this work agrees with the free-energies calculated using standard equilibrium thermodynamic integration techniques from Ref.~\cite{sandoval}, shown as orange squares here.}
  \end{figure}

  We repeated the procedure above for the bcc, fcc, and hcp structures. The system size was $18\times18\times18$ ($11,664$ atoms) for the bcc phase, $14\times14\times14$ ($10,976$ atoms) for the fcc phase, and $19\times11\times12$ ($10,032$ atoms) for the hcp phase. The reference temperature was taken at $T_0 = 100\K$ and the switching times for all thermodynamic integration methods was $t_\t{s} = 2\ns$. To verify the result of the RS simulations at temperatures $T > T_0$ we have performed extra Frenkel--Ladd simulations at $T = 400,\, 700,\, 1000,\, 1300,\, \t{and}\, 1600\K$. Figure \ref{fig:bcc_fcc} shows the free-energy difference between the bcc and fcc phases. From this figure we see that this potential predicts that the $\t{bcc} \rightarrow \t{fcc}$ transition occurs at a temperature of $T_{\t{bcc}\rightarrow \t{fcc}} = 487\K$, almost $2.5$ times lower than the experimental result of $1183\K$. We also show in Fig.~\ref{fig:bcc_fcc} that our results calculated using the nonequilibrium approach agree within the error bars with the free-energy difference calculated independently using equilibrium TI techniques \cite{sandoval}. Although the authors in Ref.\cite{sandoval} did not use the fixed center-of-mass correction or account for errors due finite-size effects, we can estimate the corrections due these two effects using Eq.~\eqref{eq:cm} and Fig.~\ref{fig:bcc_size_convergence}. Our estimate is that this error in $\Delta G$ (Fig.~\ref{fig:bcc_fcc}) could be from $\pm 0.5\,\t{meV/atom}$ to $\pm 3.4\,\t{meV/atom}$, which is smaller or of the same order of magnitude as their error bars.
 
\begin{figure}[t]
    \centering
   \includegraphics[width=0.48\textwidth]{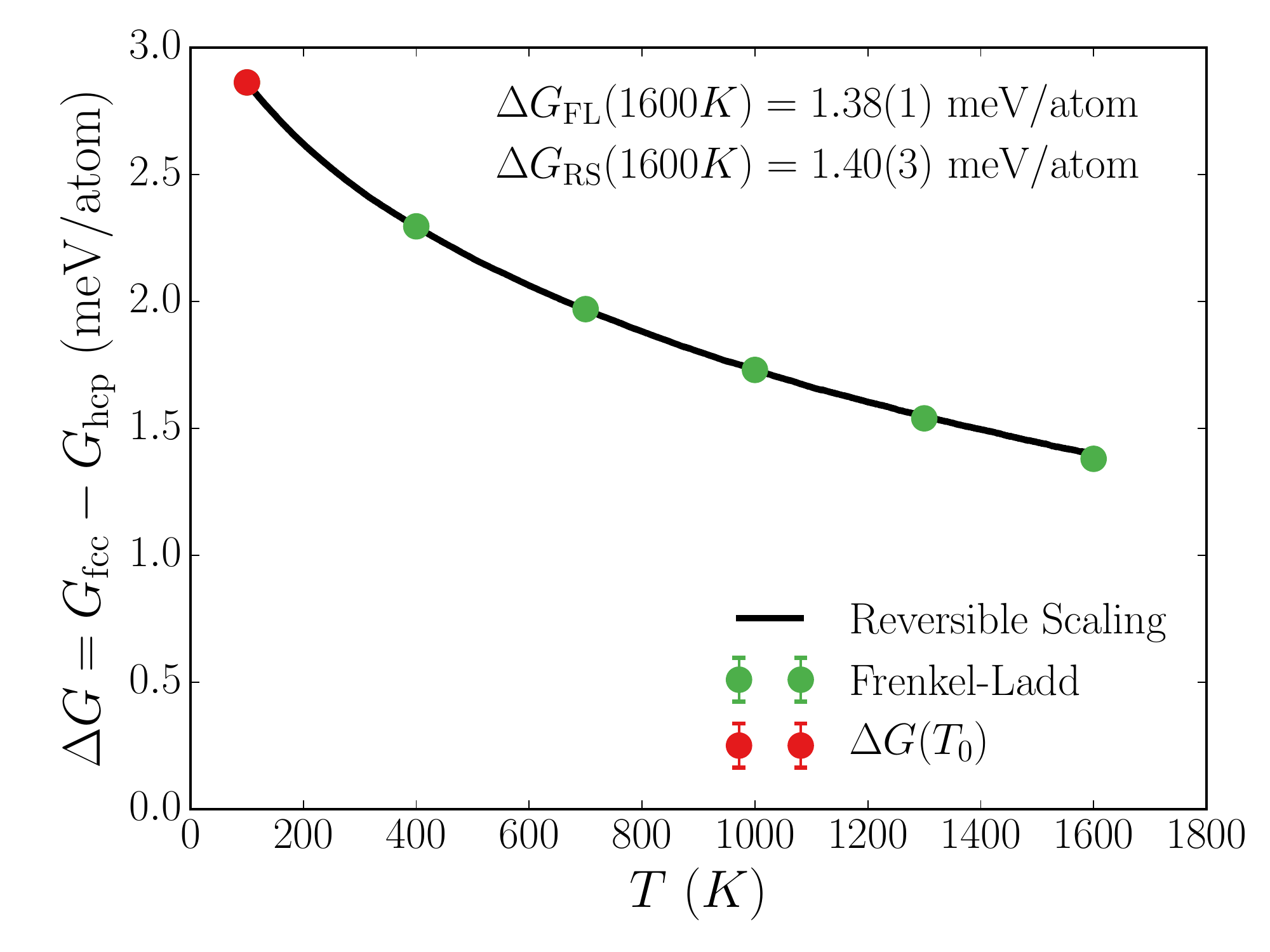}
    \caption{\label{fig:fcc_hcp}Free-energy difference between the fcc and hcp phases of iron. The hcp phase is more stable than the fcc at all temperature below the melting point. Therefore, this potential does not reproduce the correct phase stability of iron.}
  \end{figure}

  In Fig.~\ref{fig:fcc_hcp} we show free-energy difference between the fcc and the hcp phases, from which we see that hcp is more stable than fcc for the entire analyzed temperature range. Although hcp is a known stable phase of Fe at high pressures, it is not observed to be thermodynamically stable at zero pressure. Thus, the EAM potential from Ref.~\cite{eam_fe} does not predict the correct phase stability of Fe above the $\t{bcc} \rightarrow \t{hcp}$ transition temperature of $440\K$, which could have important consequences for applications of this potential in atomistic simulations above this temperature. The formalism described in this paper and its implementation in LAMMPS provides a straightforward framework for testing phase stability for other interatomic potential models for solids that may display polymorphic phase transitions below the melting point.

%% file: sec_summary.tex
In this paper a detailed account has been presented of the use of state-of-the-art nonequilibrium simulation methods to compute free energies of solids in LAMMPS using the Frenkel--Ladd \cite{frenkel_ladd} and Reversible Scaling \cite{rs} paths. The approach was demonstrated in free energy calculations for different polymorphs (fcc, bcc and hcp) in a classical potential model of iron \cite{eam_fe}. It was demonstrated that a precision of tenths of meV/atom or better can be achieved in nonequilibrium simulations for systems containing on the order of $10,000$ atoms (or more) using switching times on the order of $10^6$ MD steps. Switching times as short as $4\times10^4$ steps for the RS path or $300$ steps for the FL path are shown to be sufficient to achieve an accuracy of $1\, \t{meV/atom}$.

The implementation of the nonequilibrium simulation methods in LAMMPS presented in this manuscript provides considerable flexibility in extending the technique to study more complex properties. For example, although the applications presented above are for simple solids only, the implementation of the formalism in LAMMPS can be used to compute the vibrational contributions to the free energy of A--B compounds or alloy solid solutions with different atom types. In such applications, if the vibrational frequencies of the different atom types differ strongly, it may be advantageous to assign different spring constants, $k_\t{A}$ and $k_\t{B}$, for the species A and B. This can be accomplished by using two separate commands when calling the Frenkel--Ladd routines:
\begin{center}
    \texttt{fix f3 group\textunderscore A ti/spring \$\{k\textunderscore A\} \$\{t\textunderscore s\} \$\{t\textunderscore eq\}} \\
    \texttt{fix f4 group\textunderscore B ti/spring \$\{k\textunderscore B\} \$\{t\textunderscore s\} \$\{t\textunderscore eq\}}
\end{center}
where \texttt{group\textunderscore A} (\texttt{group\textunderscore B}) is a group containing only $A$ ($B$) atoms. The variables \texttt{t\textunderscore s} and \texttt{t\textunderscore eq} contain the switching and equilibration time respectively. In cases where the system is composed of particles with very different vibrational frequencies, the use of two different spring constants may significantly enhance the accuracy of the FL path without adding any extra computational cost. Specifically, appropriate spring constants for different atomic species can be determined by monitoring the average mean-squared displacements for each of them using Eq.~\eqref{eq:msd}. It should be noted that in applications of the formalism presented in this paper to systems with more than one chemical species, the calculated free energy includes only vibrational contributions for a given atomic configuration. The contributions from configurational entropy must be included separately through mean-field statistical-mechanical models or Monte-Carlo simulations. The same nonequilibrium methods can also be applied to compute the free energy of fluid-phase systems. Instead of using an Einstein solid as a reference system, an appropriate Frenkel--Ladd path for this case may involve a purely repulsive potential such as the inverse-power soft-sphere fluid, for which accurate virial equations of state are available \cite{rs_2}. The applicability of the RS path, on the other hand, remains precisely the same compared to the case of a solid-phase system.

The methods outlined in this paper can also be extended to systems with point and extended defects, including surfaces and interfaces, to compute the excess free energies associated with these crystal imperfections. As a specific example, in Sec.~\ref{sec:bulk} the calculation of the free energy of a simple solid composed of $N$ particles was described; the resulting free energy can be referred to as $F_\t{bulk}$. Consider now another system with $N$ particles organized in the same crystalline lattice, but now the system also has a surface of total area $A$. With the procedure presented in Sec.~\ref{sec:bulk} the free energy of this system, $F_\t{surf}$, can be computed. The presence of the surface increases the free energy of the system with respect the free energy of $F_\t{bulk}$ by $\gamma A$, where $\gamma$ is the surface free energy per unit area. Hence, the total free energy of the system is $F_\t{surf} = F_\t{bulk} + \gamma A$, and the surface free energy of the specific surface orientation considered can be computed as $\gamma = (F_\t{surf} - F_\t{bulk})/A$. Applications of this approach to the study of surface free energies will be presented in a forthcoming article, where it is demonstrated that sufficient precision can be achieved with the nonequilibrium approach to enable accurate calculations of defect free energies from this procedure, even though it involves subtraction of relatively large numbers. Note that in this example the surface could be substituted by another general interface such as a grain boundary. In such calculations, it is important that the defect remain structurally ordered at the temperatures where FL path is applied, as the presence of structural disorder that evolves during the switching (\eg, due to premelting or the dynamic formation of interfacial point defects) leads to increased irreversibility in the nonequilibrium approach. For such applications an appropriate strategy would be to apply the FL path at relatively low temperatures where the interface remains structurally ordered, and perform the RS approach to compute temperature dependences of the interfacial free energies up to higher temperatures where structural disorder may be present.

The formalism presented in this work also provides an efficient framework for performing automated calculations of the free energies of simple solids that we expect will be useful in the development and benchmarking of classical interatomic potential models. For example, scripts can be readily developed employing the commands outlined above to enable automated calculations of the free energies of typical crystal structures for a given potential model. This can be useful, to understand the bulk thermodynamic properties and phase stability predicted by the potential model prior to its application in the simulation of more complex phenomena that may be affected by these properties. One could also envision applications of the above formalism for tabulating such data for all potentials available on community repositories \cite{nist}, to guide selection of a particular potential for a given application.